\documentclass[journal,twoside,web]{ieeecolor}
\usepackage{arxiv}
\usepackage{amsmath,amssymb}
\usepackage{algorithmic}
\usepackage{graphicx}
\usepackage{color}
\usepackage{booktabs}
\usepackage{makecell}
\usepackage{multirow}
\usepackage{mathrsfs}
\usepackage{subcaption}
\usepackage{rotating}
\usepackage{balance}
\usepackage{cite}
\usepackage{ulem}
\usepackage{multicol}

\usepackage{xcolor}
\definecolor{wrong}{rgb}{.8,.349,.1}
\definecolor{right}{rgb}{.3,.7,.1}
\usepackage{newfloat}
\usepackage{listings}
\usepackage{orcidlink}

\DeclareGraphicsExtensions{.pdf,.png,.jpg,.fig,.tif}
\definecolor{color1}{rgb}{0.2353,0.2353,1}
\definecolor{color2}{rgb}{0.8000,0,0.4000}
\definecolor{color3}{rgb}{0.6980,0.4000,1}
\definecolor{color4}{rgb}{1,0.6,1}

\newcolumntype{L}[1]{>{\raggedright\arraybackslash}p{#1}}
\newcolumntype{C}[1]{>{\centering\arraybackslash}p{#1}}
\newcolumntype{R}[1]{>{\raggedleft\arraybackslash}p{#1}}

\def\BibTeX{{\rm B\kern-.05em{\sc i\kern-.025em b}\kern-.08em
    T\kern-.1667em\lower.7ex\hbox{E}\kern-.125emX}}
\begin{document}
\title{Leveraging Fixed and Dynamic Pseudo-labels for Semi-supervised Medical Image Segmentation}

\author{Suruchi Kumari$^{\orcidlink{0009-0005-1296-4207}}$, \and Pravendra Singh$^{\orcidlink{0000-0003-1001-2219}}$
\thanks{Manuscript received May 11, 2024;}
\thanks{(Corresponding author: Pravendra Singh.)}
\thanks{Suruchi Kumari and Pravendra Singh are with the Department of Computer Science and Engineering, Indian Institute of Technology Roorkee, Uttarakhand 247667, India, (e-mail: suruchi\_k@cs.iitr.ac.in; pravendra.singh@cs.iitr.ac.in).}
}

\maketitle
\begin{abstract}
Semi-supervised medical image segmentation has gained growing interest due to its ability to utilize unannotated data. The current state-of-the-art methods mostly rely on pseudo-labeling within a co-training framework. These methods depend on a single pseudo-label for training, but these labels are not as accurate as the ground truth of labeled data. Relying solely on one pseudo-label often results in suboptimal results. To this end, we propose a novel approach where multiple pseudo-labels for the same unannotated image are used to learn from the unlabeled data: the conventional fixed pseudo-label and the newly introduced dynamic pseudo-label. By incorporating multiple pseudo-labels for the same unannotated image into the co-training framework, our approach provides a more robust training approach that improves model performance and generalization capabilities. We validate our novel approach on three semi-supervised medical benchmark segmentation datasets, the Left Atrium dataset, the Pancreas-CT dataset, and the Brats-2019 dataset. Our approach significantly outperforms state-of-the-art methods over multiple medical benchmark segmentation datasets with different labeled data ratios. We also present several ablation experiments to demonstrate the effectiveness of various components used in our approach.
\end{abstract}

\section{Introduction}
\label{sec.introduction}

\IEEEPARstart{P}recise image segmentation is essential in medical image analysis because it can identify important organs or lesions within medical images. Deep neural networks (DNNs) have demonstrated remarkable results in medical image segmentation. The effectiveness of DNNs in segmentation tasks heavily relies on the accessibility of pixel-wise annotated data. However, the process of collecting annotated data presents significant challenges, especially in the medical domain \cite{shen2023survey,litjens2017survey}, where noise interference and low contrast in medical images frequently result in poor visual quality. Additionally, annotating medical images demands a higher level of professional expertise compared to natural images. The challenge becomes more apparent when dealing with volumetric segmentation tasks from medical images like computed tomography (CT) or magnetic resonance imaging (MRI); depending on the complexity of the task, it may take minutes to hours to annotate a single image \cite{wang2021annotation}. On the other hand, obtaining unlabeled data is often easier and more affordable for many tasks. Consequently, semi-supervised learning (SSL) has become increasingly popular and widely used in medical image analysis \cite{chen2022recent} in recent years. In SSL, the model is trained using both labeled and unlabeled data, with labeled data being limited in quantity compared to the abundant unlabeled data available.

The main challenge in semi-supervised medical image segmentation lies in efficiently leveraging unlabeled images. This is typically addressed by assigning pseudo semantic masks to unlabeled images and subsequently integrating both labeled and pseudo-labeled images for joint training. Many existing methods rely on self-training, where a single model creates pseudo-labels and learns iteratively in a bootstrapping manner \cite{chen2022recent}. In contrast, other approaches utilize the co-training framework \cite{qiao2018deep}, where two models are simultaneously updated. Each model is trained to predict the results of its counterpart within this framework \cite{jiao2023learning,kumari2023data}. Many state-of-the-art SSL approaches are built upon the principles of co-training \cite{sohn2020fixmatch,luo2022semi}. As pseudo-labels inevitably include noise, simply incorporating these predictions into model training is likely to trigger the confirmation bias problem \cite{arazo2020pseudo}, i.e., overfitting to the incorrect pseudo-labels \cite{ma2023enhanced}. To address these challenges, various strategies have been developed. For example, some methods heavily rely on a high-confidence threshold to calculate the unsupervised loss, discarding pseudo-labels with confidence below this threshold \cite{sohn2020fixmatch, mahmood2024splal}. While effective, one potential issue arising from solely relying on high-confidence predictions is that certain pixels might not be learned throughout the entire training process. Alternative methods aim to mitigate model uncertainty \cite{yu2019uncertainty, shi2021inconsistency}. These methods use estimated uncertainty measures to filter out unreliable predictions, keeping only those with low uncertainty for calculating the loss function.

The efficiency of the semi-supervised models heavily relies on the quality and reliability of the pseudo-labels. Semi-supervised learning can approach supervised learning if the pseudo-labels of unlabeled data are accurate enough to approximate the ground truth. Previous SSL approaches rely on a single pseudo-label to learn from unlabeled data, despite the fact that the pseudo-label is noisy and not as accurate as ground truth. While there are effective ways for determining reliable and high confidence pseudo-labels, these methods are bottlenecked by the quality of this single pseudo-label.

To address the above issues, instead of relying on a single pseudo-label for the entire training of unlabeled data, we utilize multiple pseudo-labels for the same unlabeled image to distribute the learning burden across multiple pseudo-labels. To achieve this, we propose a novel approach where multiple pseudo-labels are used to learn from the unlabeled data: the conventional fixed pseudo-label and the newly introduced dynamic pseudo-label. This additional pseudo-label introduces variations while learning from unlabeled data and makes the model more capable of learning new patterns in the data. We create dynamic pseudo-label \(\hat{y}^{u_d}\) by combining the non-overlapping region from the fixed pseudo-label \(\hat{y}^{u_f}\) and the overlapping region from the temporary pseudo-label \(\hat{y}^{u_t}\). Temporary pseudo-label is generated by cutting four additional cropped images $C = {\{c_{d_1}, c_{d_2}, c_{d_3}, c_{d_4}\}}$ with respect to the fixed cropped image $x^{u_f}$ from the original image $I$ as shown in Figure \ref{fig.shift}. The first additional cropped image $c_{d_1}$ is obtained by shifting the image in a left horizontal direction by $\sigma$ pixels with respect to fixed cropped image $x^{u_f}$. Similarly, $c_{d_2}$ is obtained by shifting the image in the right horizontal direction by $\sigma$ pixels with respect to fixed cropped image $x^{u_f}$, the same way $c_{d_3}$ and $c_{d_4}$ are obtained by shifting the image in the vertical direction by $\sigma$ pixels with respect to fixed cropped image $x^{u_f}$ as shown in Figure \ref{fig.shift}. Finally, one image out of four additional cropped images is randomly used to generate the temporary pseudo-label \(\hat{y}^{u_t}\). This process introduces randomness into the choice of pseudo-label used for unsupervised loss calculation. It is important to note that since we shifted the image to $\sigma$  pixels with respect to fixed cropped image $x^{u_f}$ to generate the temporary pseudo-label \(\hat{y}^{u_t}\), we use the $COM(.)$ operation to avoid applying loss to regions that do not overlap. This operation creates a dynamic pseudo-label \(\hat{y}^{u_d}\) by taking the non-overlapping region from the fixed pseudo-label \(\hat{y}^{u_f}\) and the overlapping region from the temporary pseudo-label \(\hat{y}^{u_t}\), as shown in Figure \ref{fig:main}. By incorporating two pseudo-labels (i.e. fixed and dynamic) in model learning, we introduce greater diversity into the co-training framework, thereby enhancing its representation learning and generalization ability.

We assess the effectiveness of our proposed approach across three medical image segmentation datasets: LA, Pancreas-CT, and Brats-2019 datasets. The experiments reveal that our simple yet highly effective approach significantly outperforms current state-of-the-art methods. The ablation study further shows the effectiveness of each proposed module in our approach.

\section{Related Works}
\label{sec.related_works}

\subsection{Semi-supervised learning}

Existing deep semi-supervised learning methods are usually categorized into (1) deep generative methods: These methods employ generative models like generative adversarial networks (GANs), variational autoencoders (VAEs), and their variations, \cite{lei2022semi, wu2022exploring} (2) pseudo-labeling methods: It involves assigning pseudo-labels to unlabeled examples using guidance from labeled examples \cite{bai2023bidirectional, su2024mutual}, (3) consistency regularization methods: It ensures that the prediction for an unlabeled example remains consistent despite various perturbations, such as noise introduction or data augmentation \cite{chen2023semi, miao2023sc, ouali2020semi}, (4) hybrid approaches: These methods combine various learning strategies including consistency regularization, pseudo-labeling, and data augmentation \cite{jiao2023learning}. 

Pseudo-labeling methods are often classified as either co-training or self-training. In self-training approaches, the algorithm starts with a small set of labeled data to make predictions. These predictions are then used to create pseudo-labels for additional unlabeled data, thus gradually enlarging the training set. Conversely, in the Co-training framework \cite{qiao2018deep}, a model is trained on a dataset that incorporates multiple perspectives or views of the data. These views usually vary but offer complementary information. These diverse views can arise from employing various data augmentation techniques, utilizing multiple modalities, implementing various network architectures, or employing different feature extraction methods. Chen et al. \cite{chen2021semi} used distinct initializations to perturb the two segmentation networks and ensure consistency when processing the same input image. Conversely, Ouali et al. \cite{ouali2020semi} make use of unlabeled data by ensuring that the predictions made by the primary decoder and the auxiliary decoders are consistent. This is accomplished by utilizing various perturbed versions of the encoder's output. Recently, Nguyen et al. \cite{nguyen2023cross} introduced a regularization method to further reinforce the smoothness assumption for segmentation tasks. In a co-training framework, model diversity is essential. Consequently, Zhenxi et al. employ the approach outlined in \cite{10.1007/978-3-031-43895-0_18} to enhance prediction diversity in consistency learning by incorporating a wider array of semantic information derived from multiple inputs.

\subsection{Semi-supervised medical image segmentation}

Advances in semi-supervised medical image segmentation research are motivated by the challenges of obtaining annotated medical image data. Several approaches employ the Mean Teacher (MT) model to improve the segmentation performance. For instance, UA-MT \cite{yu2019uncertainty} uses Monte Carlo sampling to assess the uncertainty associated with each target prediction in addition to producing target outputs. SASSNet \cite{li2020shape} introduces a shape-aware semi-supervised approach and enforces geometric shape constraints on segmentation outputs. A dual-task deep network known as DTC \cite{luo2021semi} concurrently predicts both the geometry-aware level set representation of a target and a pixel-wise segmentation map. The SS-Net \cite{wu2022exploring} employs two key techniques: inter-class separation and pixel-level smoothness. Inter-class separation promotes the convergence of class features towards their respective high-quality prototypes, thereby compacting each class distribution and facilitating clear separation between different classes. While the latter ensures that the model produces consistent results even when faced with adversarial changes. On the other hand, Several methods \cite{lei2022semi, bai2023bidirectional} consider the relationship between labeled and unlabeled data. ASE-Net \cite{lei2022semi} employs an adversarial consistency training approach with two discriminators that leverage consistency learning to derive prior relationships between annotated and unannotated data. BCP \cite{bai2023bidirectional} employs a straightforward Mean Teacher architecture to copy and paste annotated and unlabeled data bidirectionally. In a recent study, \cite{miao2023caussl} introduces CauSSL, a new causal diagram aimed at establishing a theoretical basis for widely used semi-supervised segmentation methods. This approach provides adaptability by smoothly integrating with a variety of existing SSL methods, thus boosting their overall effectiveness. To address the problem of unreliable pseudo-labeling, Su et al. \cite{su2024mutual} select higher confidence scores to indicate more reliable pseudo-labels. Subsequently, intra-class similarity is used to evaluate the reliability of these pseudo-labels.

In contrast to the aforementioned techniques, which employ a single pseudo-label, we use multiple pseudo-labels (i.e., fixed and dynamic) for the same unlabeled image. This simple yet effective approach improves the model's generalization ability and segmentation accuracy.

\begin{figure}[!t]
	\begin{center}
		\centering
		\includegraphics[width=0.35\textwidth]{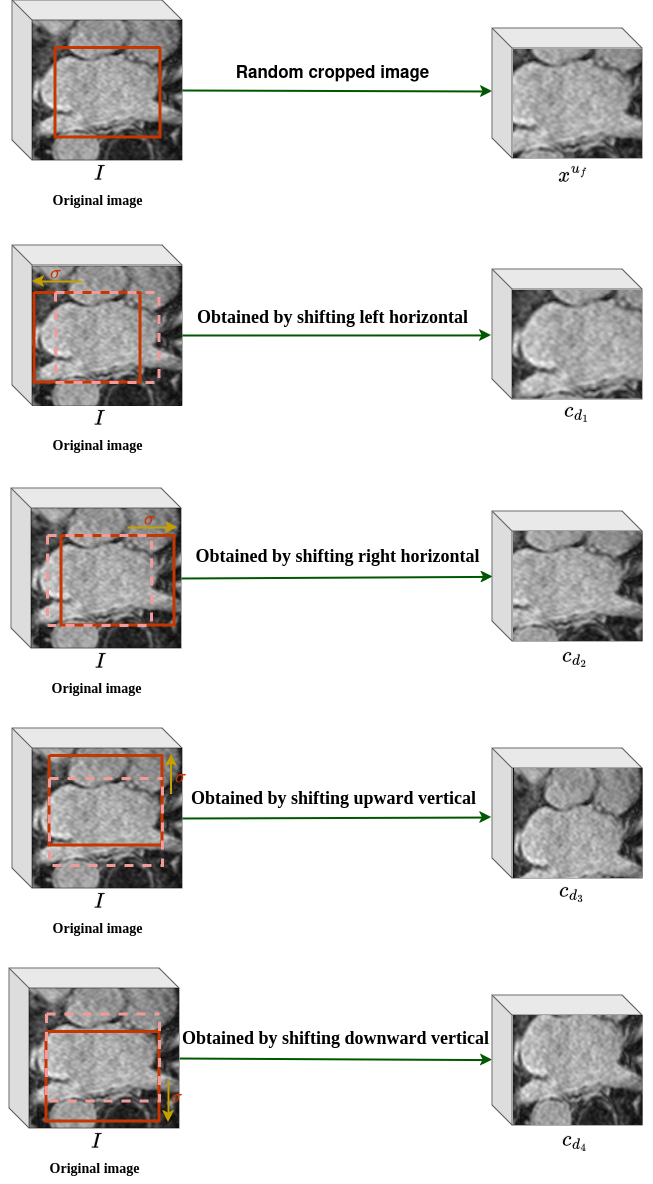}
		\centering
		\caption{The process of obtaining shifted cropped images $C = {\{c_{d_1}, c_{d_2}, c_{d_3}, c_{d_4}\}}$. Where $x^{u_f}$ represents the fixed cropped image extracted by random cropping, and $c_{d_1}, c_{d_2}, c_{d_3}, c_{d_4}$ are the cropped images extracted by applying shifting with respect to fixed cropped image $x^{u_f}$.}
		\label{fig.sl_vs_ssl}
  \label{fig.shift}
	\end{center}
\end{figure}

\section{Methodology}
\label{sec.methodology}

In a general 3D semi-supervised segmentation problem, the training data comprises $M$ labeled images and $N$ unlabeled images, denoted as, $D_l=\left\{\left(x_i, y_i \right)\right\}_{i=1}^{M}$ and $D_u=\left\{x_i \right\}_{i=1}^{N}$, respectively. Generally, $N \gg M$. Here, $x_i \subset \mathbb{R}^{W\times H\times L}$ depicts the input volume with dimensions $W\times H \times L$, and $y_i \subset \left\{ 0, 1\right\}^{H\times W\times L \times Y}$ represents the ground truth for each pixel, where $Y$ signifies the total number of visual classes. The aim is to utilize the datasets $D_l$ and $D_u$ to train a model to generate meaningful semantic predictions.

\subsection{Co-Training Framework.}
\label{sec.tnn}

Our method comprises two subnets, denoted as subnet $\mathcal{SN}_1$ and subnet $\mathcal{SN}_2$, sharing similar architectures. A batch of input data $\mathbf{D}$ comprises labeled data $D_l$ and unlabeled data $D_u$, which are processed by both subnets. The outputs comprise predictions for both labeled and unlabeled volumes. The loss function of each subnet consists of supervised and unsupervised loss components. Specifically, to learn from labeled data $D_l$, we use conventional segmentation losses, including cross-entropy loss ($L_{\text{ce}}$) and dice loss ($L_{\text{dice}}$). Our supervised loss for both subnets is formulated as follows:

\begin{equation}\label{eq_sup_single}
\begin{split}
    \mathcal{L}_{sup,\mathcal{SN}_1}^l &= \frac{1}{M} \sum_{m=1}^{M} \frac{1}{W \times H \times L} \Bigg(\sum_{n=0}^{W \times H \times L} \bigg(\ell_{ce}(\widetilde{y}_{mn,\mathcal{SN}_1}^{l}, y_{mn}^{l}) \\
    &\quad + \ell_{dice}(\widetilde{y}_{mn,\mathcal{SN}_1}^{l}, y_{mn}^{l})\bigg)\Bigg)
\end{split}
\end{equation}

\begin{equation}\label{eq_sup_single}
\begin{split}
    \mathcal{L}_{sup,\mathcal{SN}_2}^l &= \frac{1}{M} \sum_{m=1}^{M} \frac{1}{W \times H \times L} \Bigg(\sum_{n=0}^{W \times H \times L} \bigg(\ell_{ce}(\widetilde{y}_{mn,\mathcal{SN}_2}^{l}, y_{mn}^{l}) \\
    &\quad + \ell_{dice}(\widetilde{y}_{mn,\mathcal{SN}_2}^{l}, y_{mn}^{l})\bigg)\Bigg)
\end{split}
\end{equation}

Here $y_{mn}^l$ represents the ground-truth label for $n$-th pixel in the $m$-th labeled image. Similarly, $\widetilde{y}_{mn,\mathcal{SN}_1}^{l}$ and $\widetilde{y}_{mn,\mathcal{SN}_2}^{l}$ represent the model predictions for the $n$-th pixel in the $m$-th labeled image for subnet 1 ($\mathcal{SN}_1$) and 2 ($\mathcal{SN}_2$) respectively.

To learn from unlabeled data $D_u$, both subnet $\mathcal{SN}_1$ and subnet $\mathcal{SN}_2$ utilize each other's pseudo-labels via unsupervised loss. A subnet $\mathcal{SN}_1$ prediction for unlabeled image is represented by $\widetilde{y}_{mn,\mathcal{SN}_1}^{u}$ and the pseudo-label produced by it can be expressed as $\hat{y}_{mn, \mathcal{SN}_1}^{u} = \mathcal{S}(\widetilde{y}_{mn,\mathcal{SN}_1}^{u})$, where $\mathcal{S}$ is the sharpening function \cite{xie2020unsupervised}. The sharpening function is utilized to transform the network predictions into soft pseudo-labels. Similarly, subnet $\mathcal{SN}_2$ prediction and pseudo-labels is denoted by $\widetilde{y}_{mn,\mathcal{SN}_2}^{u}$ and $\hat{y}_{mn, \mathcal{SN}_2}^{u}$ respectively. We utilize the mean squared error (MSE) loss as an unsupervised loss function. The total loss for each subnet is calculated separately, as shown in Equations \ref{eq_total1} and \ref{eq_total2}.

As previously mentioned, depending on a single pseudo-label leads to suboptimal results. To overcome this, we generate multiple pseudo-labels for the same unlabeled data. Dynamic pseudo-label is obtained by cutting additional cropped images $C = {\{c_{d_1}, c_{d_2}, c_{d_3}, c_{d_4}\}}$ along with the fix cropped image $x^{u_f}$ from the original image $I$ as shown in Figure \ref{fig.shift}. From the four cropped images, we select one randomly and then we construct the pseudo-label for that image and finally utilize it in the dynamic unsupervised loss.

\begin{figure*}[!t]
	\begin{center}
		\includegraphics[width=0.99\textwidth]{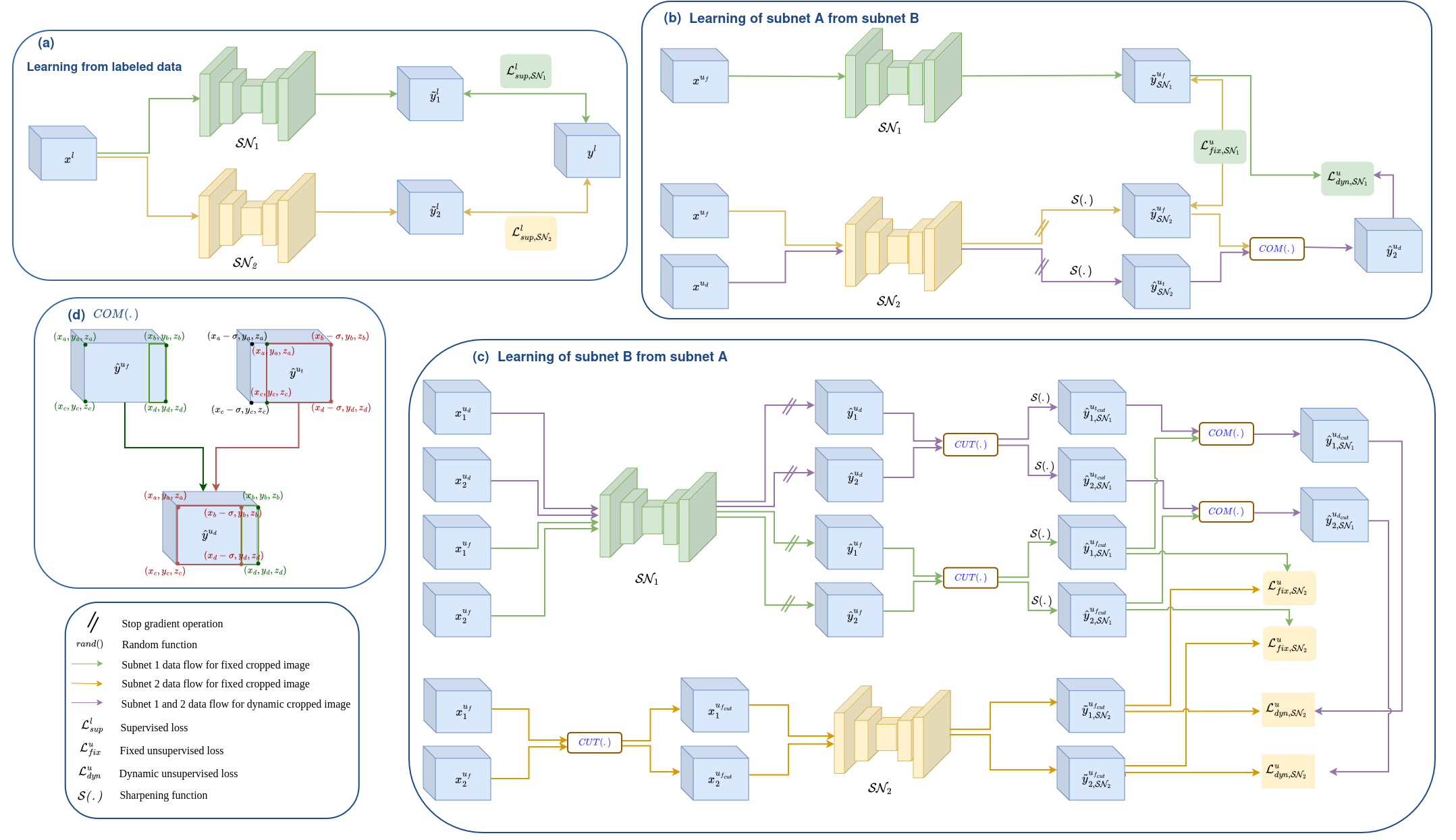}
		\caption{
			An illustration of our fixed and dynamic pseudo-labeling strategy. (a) To learn from labeled data, supervised loss $\mathcal{L}_{sup}^l$ is applied between the predictions of both networks and the ground truth. (b) When subnet A ($\mathcal{SN}_1$) learns from subnet B ($\mathcal{SN}_2$), both subnet receives the original image and $\mathcal{L}_{fix}^u$ is applied between the prediction of  $\mathcal{SN}_1$ and fixed pseudo-label of  $\mathcal{SN}_2$. Similarly, $\mathcal{L}_{dyn}^u$ is applied between the prediction of  $\mathcal{SN}_1$ and dynamic pseudo-label of  $\mathcal{SN}_2$. (c) When $\mathcal{SN}_2$ learns from $\mathcal{SN}_1$, $\mathcal{SN}_1$ receives the original image while $\mathcal{SN}_2$ receives the cut-mix image. The same fixed and dynamic loss is then applied between both subnets. (d) $COM(.)$ operation creates a dynamic pseudo-label \(\hat{y}^{u_d}\) by taking the non-overlapping region from the fixed pseudo-label \(\hat{y}^{u_f}\) and the overlapping region from the temporary pseudo-label \(\hat{y}^{u_t}\).}

    \label{fig:main}
	\end{center}
\end{figure*}

\subsection{Dynamic Pseudo-label}

Suppose our original image $I$ has dimensions \(H \times W \times L\). Following various methods {\cite{xie2023deep,shi2021inconsistency,wu2021semi,wu2022exploring}, we extract a cropped image from the original image, denoted as $x^{u_f}$, which has dimensions \(H' \times W' \times L'\), where $H>H', W>W'$ and $L>L'$. In our method, besides the image $x^{u_f}$, we also crop four additional images from the original unlabeled image, with each being relative to the image $x^{u_f}$  by \(\sigma\) pixels position in either horizontal or vertical directions. If the location of $x^{u_f}$ is represented by Cartesian coordinates \((x_a, y_a, z_a)\), \((x_b, y_b, z_b)\), \((x_c, y_c, z_c)\), and \((x_d, y_d, z_d)\), then we extract the other four images by shifting \(\sigma\) pixels in four directions (left horizontal, right horizontal, upward vertical, and downward vertical), which is formalized as follows:

\begin{align*}
c_{d_1} &= CROP(I, (x_a - \sigma, y_a, z_a), (x_b - \sigma, y_b, z_b), \\
    &\quad\quad\quad\quad\quad\quad\quad\quad\quad\quad\quad (x_c - \sigma, y_c, z_c), (x_d - \sigma, y_d, z_d)) \\
c_{d_2} &= CROP(I, (x_a + \sigma, y_a, z_a), (x_b + \sigma, y_b, z_b), \\
    &\quad\quad\quad\quad\quad\quad\quad\quad\quad\quad\quad (x_c + \sigma, y_c, z_c), (x_d + \sigma, y_d, z_d)) \\
c_{d_3} &= CROP(I, (x_a, y_a - \sigma, z_a), (x_b, y_b - \sigma, z_b), \\
    &\quad\quad\quad\quad\quad\quad\quad\quad\quad\quad\quad (x_c, y_c - \sigma, z_c), (x_d, y_d - \sigma, z_d)) \\
c_{d_4} &= CROP(I,(x_a, y_a + \sigma, z_a), (x_b, y_b + \sigma, z_b), \\
    &\quad\quad\quad\quad\quad\quad\quad\quad\quad\quad\quad (x_c, y_c + \sigma, z_c), (x_d, y_d + \sigma, z_d))
\end{align*}

Here, $CROP(.)$ is the operation that crops an image of the same size as the specified coordinates. We obtain dynamic cropped image ${x}^{u_d}$ by randomly selecting one cropped image from the given four additional cropped images \(c_{d_1}\), \(c_{d_2}\), \(c_{d_3}\), and \(c_{d_4}\), as follows:

\[{x}^{u_d} = RAND (c_{d_1}, c_{d_2}, c_{d_3}, c_{d_4} )\]

\noindent Here, the term $RAND$ represents a random operation that selects one of the provided cropped images. This selection introduces randomness into the choice of pseudo-label used for unsupervised loss calculation.

 Since \(c_{d_1}\), \(c_{d_2}\), \(c_{d_3}\), and \(c_{d_4}\) is shifted by $\sigma$ pixels, the overlapping region between the unlabeled prediction \(\Tilde{y}^{u_f}\) from one model and the temporary pseudo-label \(\hat{y}^{u_t}\) from another model is different. Hence, to avoid applying loss to the regions that are uncommon, we use the $COM(.)$ operation. This operation generates a dynamic pseudo-label \(\hat{y}^{u_d}\) by concatenating the overlapping region from the temporary pseudo-label \(\hat{y}^{u_t}\) and the non-overlapping regions from the fixed pseudo-label \(\hat{y}^{u_f}\), as depicted in Figure \ref{fig:main}.

\subsubsection{Fixed and dynamic unsupervised loss}

To train the subnetworks, we pass the image \(x^{u_f}\) to one subnet to obtain the prediction \(\Tilde{y}^{u_f}\), while we pass \(x^{u_f}\) and ${x}^{u_d}$ to another subnet to obtain the fixed \(\hat{y}^{u_f}\) and temporary pseudo-label \(\hat{y}^{u_t}\), as shown in Figure \ref{fig:main}. As mentioned above, we use the $COM(.)$ operation to generate the dynamic pseudo label \(\hat{y}^{u_d}\). Further, we apply the conventional MSE loss between the prediction of one model \(\Tilde{y}^{u_f}\) and fixed pseudo-label of another model \(\hat{y}^{u_f}\), we call this a fixed pseudo-label loss, as shown in Equations \ref{eq_fix_1} and \ref{eq_fix_2}. Similarly, for our dynamic loss, we utilize the same MSE loss between the prediction of one model \(\Tilde{y}^{u_f}\) and dynamic pseudo-label of another model \(\hat{y}^{u_d}\), as shown in Equations \ref{eq_dyn_1} and \ref{eq_dyn_2}. 

\begin{equation}\label{eq_fix_1}
\begin{split}
    \mathcal{L}_{fix,\mathcal{SN}_1}^u &= \frac{1}{N} \sum_{m=1}^{N} \frac{1}{W \times H \times L} \Bigg(\sum_{n=0}^{W \times H \times L} \\
     &\quad \ell_{mse}(\widetilde{y}_{mn,\mathcal{SN}_1}^{u_f}, \hat{y}_{mn,\mathcal{SN}_2}^{u_f})\Bigg)
\end{split}
\end{equation}

\begin{equation}\label{eq_dyn_1}
\begin{split}
     \mathcal{L}_{dyn,\mathcal{SN}_1}^u &= \frac{1}{N} \sum_{m=1}^{N} \frac{1}{W \times H \times L} \Bigg(\sum_{n=0}^{W \times H \times L} \\
     &\quad \ell_{mse}(\widetilde{y}_{mn,\mathcal{SN}_1}^{u_f}, \hat{y}_{mn,\mathcal{SN}_2}^{u_d})\Bigg)
\end{split}
\end{equation}

\begin{equation}\label{eq_fix_2}
\begin{split}
   \mathcal{L}_{fix,\mathcal{SN}_2}^u  &= \frac{1}{N} \sum_{m=1}^{N} \frac{1}{W \times H \times L} \Bigg(\sum_{n=0}^{W \times H \times L} \\
   &\quad \ell_{mse}(\widetilde{y}_{mn,\mathcal{SN}_2}^{u_{f_{cut}}}, \hat{y}_{mn,\mathcal{SN}_1}^{u_{f_{cut}}})\Bigg)
\end{split}
\end{equation}

\begin{equation}\label{eq_dyn_2}
\begin{split}
   \mathcal{L}_{dyn,\mathcal{SN}_2}^u  &= \frac{1}{N} \sum_{m=1}^{N} \frac{1}{W \times H \times L} \Bigg(\sum_{n=0}^{W \times H \times L} \\
   &\quad \ell_{mse}(\widetilde{y}_{mn,\mathcal{SN}_2}^{u_{f_{cut}}}, \hat{y}_{mn,\mathcal{SN}_1}^{u_{d_{cut}}})\Bigg)
\end{split}
\end{equation}

So, our total loss for subnet 1 is formulated as follows:

\begin{equation}\label{eq_total1}
   \mathcal{L}_{\mathcal{SN}_1} = \alpha \mathcal{L}_{sup,\mathcal{SN}_1}^l +  \mathcal{L}_{fix,\mathcal{SN}_1}^u + \beta \mathcal{L}_{dyn,\mathcal{SN}_1}^u 
\end{equation}

Similarly, for subnet 2, our total loss is formulated as follows:

\begin{equation}\label{eq_total2}
   \mathcal{L}_{\mathcal{SN}_2} = \alpha \mathcal{L}_{sup,\mathcal{SN}_2}^l + \mathcal{L}_{fix,\mathcal{SN}_2}^u + \beta \mathcal{L}_{dyn,\mathcal{SN}_2}^u 
\end{equation}

Here, the $\alpha$ hyper-parameter is used to control the contribution of the supervised loss, while the $\beta$ hyper-parameter is used to control the contribution of the unsupervised dynamic loss.

\begin{table}[t] \small 
\setlength{\tabcolsep}{0.9mm}{
\resizebox{1\linewidth}{!}{
\begin{tabular}{c|cc|llll}
\hline
\multicolumn{1}{c|}{\multirow{2}{*}{Method}} & \multicolumn{2}{c|}{Scans used}  & \multicolumn{4}{c}{Metrics} \\ 
\cline{2-7} \multicolumn{1}{c|}{} & \multicolumn{1}{l}{Labeled} & \multicolumn{1}{l|}{Unlabeled} & 
Dice$\uparrow$ & Jaccard$\uparrow$ & 95HD$\downarrow$ & ASD$\downarrow$ \\ \hline
\multicolumn{1}{c|}{V-Net} &\multicolumn{1}{c}{4(5\%)} &\multicolumn{1}{c|}{0} &52.55 &39.60 &47.05 &9.87 \\ \hline

UA-MT \cite{yu2019uncertainty}  & \multirow{7}{*}{4(5\%)} & \multirow{7}{*}{76(95\%)} 
& 82.26 & 70.98 & 13.71 & 3.82 \\
URPC \cite{luo2021efficient}  & & & 82.48 & 71.35 & 14.65 & 3.65 \\
DTC \cite{luo2021semi}  & & & 81.25 & 69.33 & 14.90 & 3.99 \\
SASSNet \cite{li2020shape} &  &  & 81.60 & 69.63 & 16.16 & 3.58 \\

MC-Net \cite{wu2021semi}  & & & 83.59 & 72.36 & 14.07 & 2.70  \\
SS-Net \cite{wu2022exploring}  & & & 86.33 & 76.15 & 9.97 & 2.31 \\ 
ACTION++ \cite{you2023action++}  & & & 87.8 & NA & NA & 2.09   \\ 
BCP \cite{bai2023bidirectional}  & & & 88.02 & 78.72 & 7.90 & 2.15 \\ 
Ours & & & \textbf{89.55}{\color{right}\textbf{\scriptsize{$\uparrow$}1.53}}&\textbf{81.18}{\color{right}\textbf{\scriptsize{$\uparrow$}2.46}}&\textbf{5.48}{\color{right}\textbf{\scriptsize{$\downarrow$}2.42}}&\textbf{1.99}{\color{right}\textbf{\scriptsize{$\downarrow$}0.16}} \\                 \hline

\end{tabular}}
}
\caption{Comparison of the proposed approach with other approaches on the LA dataset using 5 percent labeled data. Improvements relative to the second-best results are \color{right}{highlighted}.}

\label{tab:LA_4}
\end{table}

\begin{table}[t] \small 
\setlength{\tabcolsep}{0.9mm}{
\resizebox{1\linewidth}{!}{
\begin{tabular}{c|cc|llll}
\hline
\multicolumn{1}{c|}{\multirow{2}{*}{Method}} & \multicolumn{2}{c|}{Scans used}  & \multicolumn{4}{c}{Metrics} \\ 
\cline{2-7} \multicolumn{1}{c|}{} & \multicolumn{1}{l}{Labeled} & \multicolumn{1}{l|}{Unlabeled} & 
Dice$\uparrow$ & Jaccard$\uparrow$ & 95HD$\downarrow$ & ASD$\downarrow$ \\ \hline

\multicolumn{1}{c|}{V-Net} &\multicolumn{1}{c}{8(10\%)} &\multicolumn{1}{c|}{0} &82.74 &71.72 &13.35 &3.26 \\ \hline

UA-MT \cite{yu2019uncertainty}  & \multirow{10}{*}{8(10\%)} & \multirow{10}{*}{72(90\%)} 
& 87.79 & 78.39 & 8.68 & 2.12 \\
URPC \cite{luo2021efficient}  & & & 86.92 & 77.03 & 11.13 & 2.28  \\

DTC \cite{luo2021semi}  & & & 87.51 & 78.17 & 8.23 & 2.36  \\
SASSNet \cite{li2020shape}  & & & 87.54 & 78.05 & 9.84 & 2.59 \\ 
MC-Net \cite{wu2021semi} & & & 87.62 & 78.25 & 10.03 & 1.82   \\
SS-Net \cite{wu2022exploring}  & & & 88.55 & 79.62 & 7.49 & 1.90 \\
Simcvd \cite{you2022simcvd} & & & 89.03 & 80.34 & 8.34 & 2.59 \\
BCP \cite{bai2023bidirectional}  & & & 89.62 & 81.31 & 6.81 & 1.76 \\
ACTION++ \cite{you2023action++}  & & & 89.9 & NA & NA & 1.74   \\ 
DMD \cite{xie2023deep}  & & & 89.70 &  81.42 &  6.88 & 1.78 \\
MLRPL \cite{su2024mutual}  & & & 89.86 & 81.68 & 6.91 & 1.85 \\
Ours & & & \textbf{91.07}{$\color{right}\textbf{\scriptsize{$\uparrow$}1.17}$}&\textbf{83.67}{\color{right}\textbf{\scriptsize{$\uparrow$}1.99}}&\textbf{4.96}{\color{right}\textbf{\scriptsize{$\downarrow$}1.85}}&\textbf{1.65}{\color{right}\textbf{\scriptsize{$\downarrow$}0.09}}\\ \hline

\end{tabular}}
}
\caption{Comparison of the proposed approach with other approaches on the LA dataset using 10 percent labeled data. Improvements relative to the second-best results are \color{right}{highlighted}.}

\label{tab:LA_8}
\end{table}

\begin{table}[t] \small 
\setlength{\tabcolsep}{0.5mm}{
\resizebox{1\linewidth}{!}{
\begin{tabular}{c|cc|llll}
\hline
\multicolumn{1}{c|}{\multirow{2}{*}{Method}} & \multicolumn{2}{c|}{Scans used}  & \multicolumn{4}{c}{Metrics} \\ 
\cline{2-7} \multicolumn{1}{c|}{} & \multicolumn{1}{l}{Labeled} & \multicolumn{1}{l|}{Unlabeled} & 
Dice$\uparrow$ & Jaccard$\uparrow$ & 95HD$\downarrow$ & ASD$\downarrow$ \\ \hline
\multicolumn{1}{c|}{V-Net} &\multicolumn{1}{c}{6(10\%)} &\multicolumn{1}{c|}{0} & 54.94 & 40.87 & 47.48 & 17.43 \\ 
\multicolumn{1}{c|}{V-Net} &\multicolumn{1}{c}{12(20\%)} &\multicolumn{1}{c|}{0} & 71.52 & 57.68 & 18.12 & 5.41 \\ \hline

UA-MT \cite{yu2019uncertainty}  & \multirow{10}{*}{6(10\%)} & \multirow{10}{*}{56(90\%)} & 66.44 & 52.02 & 17.04 & 3.03 \\
SS-Net \cite{wu2022exploring} & & & 65.51 & 51.09 & 18.13 & 3.44 \\ 
DTC \cite{luo2021semi}  & & & 66.58 & 51.79 & 15.46 & 4.16 \\
SASSNet \cite{li2020shape}  &  &  & 68.97 & 54.29 & 18.83 & 1.96   \\ 

URPC \cite{luo2021efficient}  & & & 73.53 & 59.44 & 22.57 & 7.85 \\ 
MC-Net+ \cite{wu2022mutual}  & & & 70.00 & 55.66 & 16.03 & 3.87  \\
SC-SSL \cite{miao2023sc}  & & & 70.09 & 55.52 & 12.88 & 1.97  \\
MCCauSSL \cite{miao2023caussl} & & & 72.89 & 58.06 & 14.19 & 4.37 \\
MLRPL \cite{su2024mutual}  & & &  75.93	& 62.12	&9.07	&1.54\\
Ours & & & \textbf{79.06}{\color{right}\textbf{\scriptsize{$\uparrow$}3.13}}&\textbf{66.04}{\color{right}\textbf{\scriptsize{$\uparrow$}3.92}}&\textbf{7.70}{\color{right}\textbf{\scriptsize{$\downarrow$}1.37}}&\textbf{1.50}{\color{right}\textbf{\scriptsize{$\downarrow$}0.04}} \\ \hline
UA-MT \cite{yu2019uncertainty}  & \multirow{10}{*}{12(20\%)} & \multirow{10}{*}{50(80\%)} 
& 76.10 & 62.62 & 10.84 & 2.43 \\
SS-Net \cite{wu2022exploring}& & & 76.20 & 63.00 & 10.65 & 2.68 \\
DTC \cite{luo2021semi}  & & & 76.27 & 62.82 & 8.70 & 2.20  \\
SASSNet \cite{li2020shape}  & & & 76.39 & 63.17 & 11.06 & 1.42 \\ 

URPC \cite{luo2021efficient}  & & & 80.02 & 67.30 & 8.51 & 1.98  \\ 
MC-Net+ \cite{wu2022mutual}  & & & 79.37 & 66.83 & 8.52 & 1.72   \\
SC-SSL \cite{miao2023sc}  & & & 80.76 & 68.17 & 6.79 & 1.73  \\
 MCCauSSL \cite{miao2023caussl} & & & 80.92 & 68.26 & 8.11 & 1.53 \\
 MLRPL \cite{su2024mutual}  & & & 81.53 & 69.35 & 6.81 & 1.33\\
Ours & & & \textbf{83.71}{$\color{right}\textbf{\scriptsize{$\uparrow$}2.18}$}&\textbf{72.19}{\color{right}\textbf{\scriptsize{$\uparrow$}2.84}}&\textbf{4.87}{\color{right}\textbf{\scriptsize{$\downarrow$}1.94}}&\textbf{1.22}{\color{right}\textbf{\scriptsize{$\downarrow$}0.11}}\\ \hline
\end{tabular}}
}
\caption{Comparison with other methods on pancreas dataset using 10 percent and 20 percent labeled data. Improvements relative to the second-best results are \color{right}{highlighted}.}
\label{tab:pancreas}
\end{table}

\begin{table}[t] \small 
\setlength{\tabcolsep}{0.7mm}{
\resizebox{1\linewidth}{!}{
\begin{tabular}{c|cc|llll}
\hline
\multicolumn{1}{c|}{\multirow{2}{*}{Method}} & \multicolumn{2}{c|}{Scans used}  & \multicolumn{4}{c}{Metrics} \\ 
\cline{2-7} \multicolumn{1}{c|}{} & \multicolumn{1}{l}{Labeled} & \multicolumn{1}{l|}{Unlabeled} & 
Dice$\uparrow$ & Jaccard$\uparrow$ & 95HD$\downarrow$ & ASD$\downarrow$ \\ \hline
\multicolumn{1}{c|}{U-Net} &\multicolumn{1}{c}{25(10\%)} &\multicolumn{1}{c|}{0} & 74.43 & 61.86 & 37.11 & 2.79 \\ 
\multicolumn{1}{c|}{U-Net} &\multicolumn{1}{c}{50(20\%)} &\multicolumn{1}{c|}{0} & 80.16 & 71.55 & 22.68 & 3.43 \\ \hline

DAN \cite{zhang2017deep}  & \multirow{9}{*}{25(10\%)} & \multirow{9}{*}{225(90\%)} & 81.71 & 71.43  & 15.15 & 2.32 \\
CPS \cite{chen2021semi} &  &  & 82.52 & 72.66 & 13.08 & 2.66 \\
EM \cite{vu2019advent} &  &  & 82.27 & 72.15  & 11.98 & 2.42 \\
DTC \cite{luo2021semi} &  &  & 81.75 & 71.63  & 15.73 & 2.56 \\
CCT \cite{ouali2020semi} &  &  & 82.15 & 71.59  & 16.57 & 2.11 \\
URPC \cite{luo2021efficient} &  &  & 82.59 & 72.11  & 13.88 & 3.72 \\
CPCL \cite{xu2022all} &  & & 83.36 & 73.23 & 11.74 & 1.99 \\
AC-MT \cite{xu2023ambiguity} &  &  & 83.77 & 73.96  & 11.37 & 1.93 \\
MLRPL \cite{su2024mutual}  & & &  84.29 & 74.74 & 9.57 & 2.55 \\ 
Ours & & & \textbf{85.71}{\color{right}\textbf{\scriptsize{$\uparrow$}1.42}}&\textbf{76.35}{\color{right}\textbf{\scriptsize{$\uparrow$}1.61}}&\textbf{8.15}{\color{right}\textbf{\scriptsize{$\downarrow$}1.42}}&\textbf{1.83}{\color{right}\textbf{\scriptsize{$\downarrow$}0.1}} \\ \hline
DAN \cite{zhang2017deep} & \multirow{9}{*}{50(20\%)} & \multirow{9}{*}{200(80\%)} 
 & 83.31 & 73.53  & 10.86 & 2.23 \\

CPS \cite{chen2021semi} &  &  & 84.01 & 74.02 & 12.16  & 2.18 \\
EM \cite{vu2019advent} &  &  & 82.91 & 73.15  & 10.52 & 2.48 \\
DTC \cite{luo2021semi} &  &  & 83.43 & 73.56  & 14.77 & 2.34 \\
CCT \cite{ouali2020semi} &  &  & 82.53 & 72.36  & 15.87 & 2.21\\
URPC \cite{luo2021efficient} &  &  & 82.93 & 72.57  & 15.93 & 4.19 \\
CPCL \cite{xu2022all} &  &  & 83.48 & 74.08  & 9.53 & 2.08 \\
AC-MT \cite{xu2023ambiguity} &  &  & 84.63 & 74.39  & 9.50 & 2.11 \\
MLRPL \cite{su2024mutual} &  &  & 85.47 & 76.32  & 7.76 & 2.00 \\ 
Ours & & & \textbf{86.65}{$\color{right}\textbf{\scriptsize{$\uparrow$}1.18}$}&\textbf{77.72}{\color{right}\textbf{\scriptsize{$\uparrow$}1.4}}&\textbf{7.32}{\color{right}\textbf{\scriptsize{$\downarrow$}0.44}}&\textbf{1.70}{\color{right}\textbf{\scriptsize{$\downarrow$}0.3}}\\ \hline
\end{tabular}}
}
\caption{Comparison of the proposed approach with other approaches on the Brats-19 dataset using 10 percent and 20 percent labeled data. Improvements relative to the second-best results are \color{right}{highlighted}.}
\label{tab:brats2019}
\end{table}

\begin{figure}[h]
	\begin{center}
		\centering
		\includegraphics[width=0.4\textwidth]{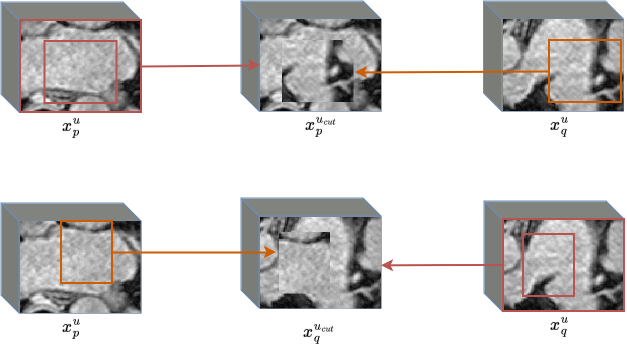}
		\centering
		\caption{An illustration of $CUT(.)$ operation. Where $x^{u_{cut}}_p$ and $x^{u_{cut}}_q$ are the images obtained after applying the $CUT(.)$ operation. $x^u_{p}$ and $x^u_{q}$ are two different unlabeled images from a batch.}
		\label{fig:cut}
	\end{center}
\end{figure}

\begin{figure*}[h!]
	\begin{center}
		\centering
		\includegraphics[width=.8\textwidth]{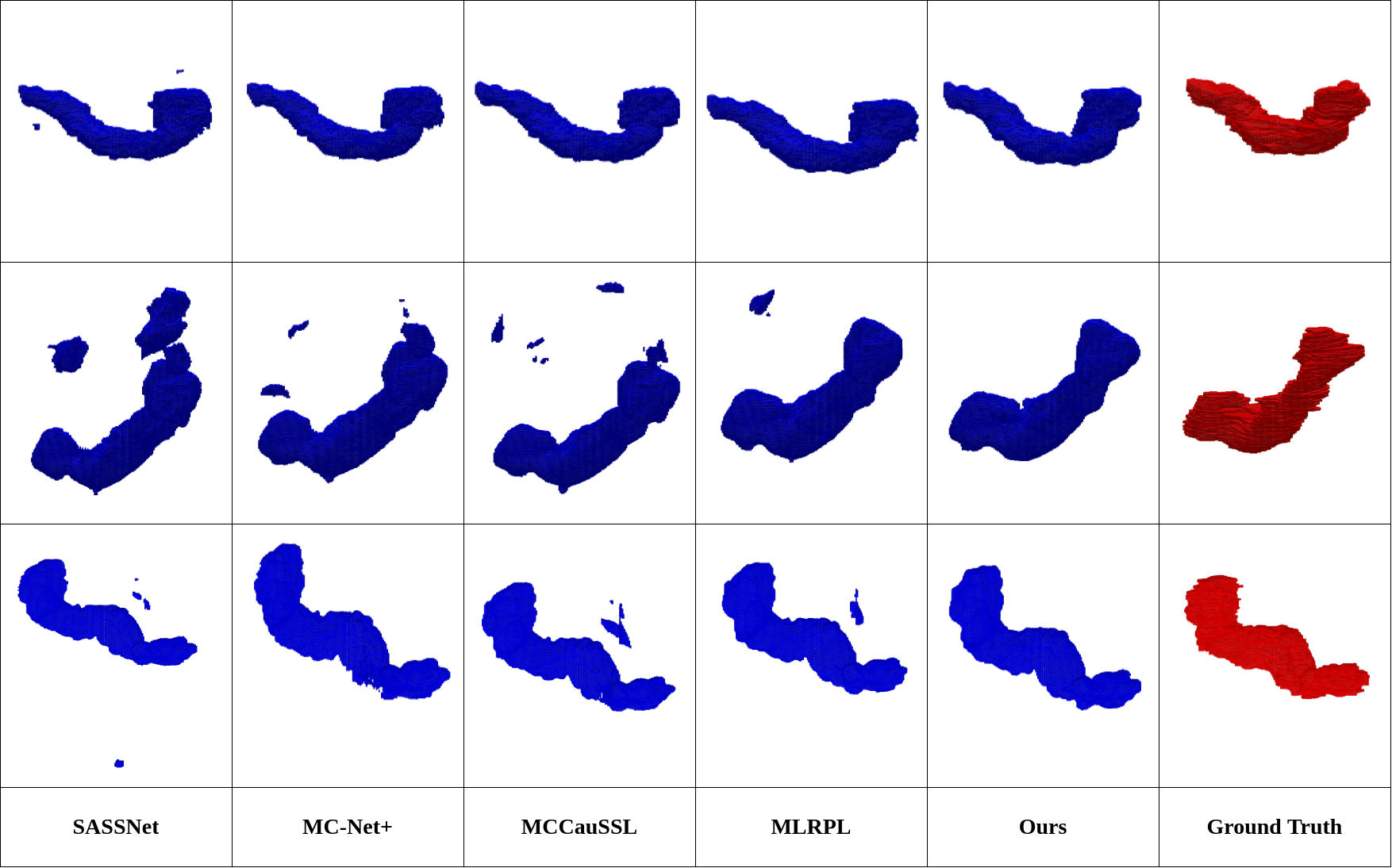}
		\centering
		\caption{Visualization results of different semi-supervised segmentation techniques are depicted, utilizing 20\% labeled data alongside ground truth on the pancreas dataset.}
		\label{fig:pancreas}
	\end{center}
\end{figure*}

\begin{table}[t]
\renewcommand\arraystretch{0.99}
\resizebox{1\linewidth}{!}{
\begin{tabular}{cccc|cccc}
\hline
$\mathcal{L}_{sup}^l$ & $\mathcal{L}_{fix}^u$ & Cut-mix & $\mathcal{L}_{dyn}^u$ & Dice
$\uparrow$ & Jaccard $\uparrow$ & 95HD $\downarrow$ & ASD $\downarrow$ \\ \hline
$\checkmark$ & $\times$  & $\times$   &  $\times$  & 71.52 & 57.68 & 18.12 & 5.41   \\
$\checkmark$ &  $\checkmark$  &  $\checkmark$   &  $\times$   & 81.28    & 68.82 & 10.13    & 2.67   \\
$\checkmark$ & $\checkmark$   &  $\times$   &   $\checkmark$  & 82.99 & 71.26 & 5.82 & 1.35   \\
$\checkmark$ &  $\times$ & $\checkmark$   & $\checkmark$     & 83.40 & 71.80 & 5.47 & 1.91  \\
$\checkmark$ & $\checkmark$   & $\checkmark$   & $\checkmark$   & \textbf{83.71} & \textbf{72.19}& \textbf{4.87} & \textbf{1.22}   \\ \hline
\end{tabular}
}
\caption{Results for ablation experiments which are conducted on the pancreas dataset using 20 \% labeled data, with a fixed value of $\beta=4$ maintained across all experiments.}
\label{tab_ablation1}
\vspace{-0.8em}
\end{table}

\begin{table}[t]\small
\renewcommand\arraystretch{0.93}
\resizebox{1\linewidth}{!}
{
\begin{tabular}{c|cc|cccc}
\hline
\multirow{2}{*}{$\beta$} & \multicolumn{2}{c|}{Scans used}                  & \multicolumn{4}{c}{Metrics} \\ \cline{2-7} 
                            & Labeled                & Unlabeled  & Dice$\uparrow$  & Jaccard$\uparrow$  & 95HD$\downarrow$ & ASD$\downarrow$ \\ \hline
1                      & \multirow{5}{*}{12(20\%)}  & \multirow{5}{*}{50(80\%)} & 83.30    & 71.62       & 7.04    & 1.91   \\
2           &      &  & 83.44    & 71.85       & 5.25    & 1.57   \\
3     &  &     & 83.54 & 72.00 & 5.06 & 1.29  \\ 
4                     &                        &                         & \textbf{83.71}   & \textbf{72.19}     & 5.47   & \textbf{1.22}   \\


5     &    &     & 83.67    & 72.21       & \textbf{4.71}   & 1.35   \\

\hline

\end{tabular}
}
\caption{Results for ablation experiments using different values of $\beta$ on pancreas dataset with 20 \% labeled data.}
\label{tab_ablation2}
\vspace{-0.8em}
\end{table}

\begin{figure}[h]
	\begin{center}
		\centering
		\includegraphics[width=0.42\textwidth]{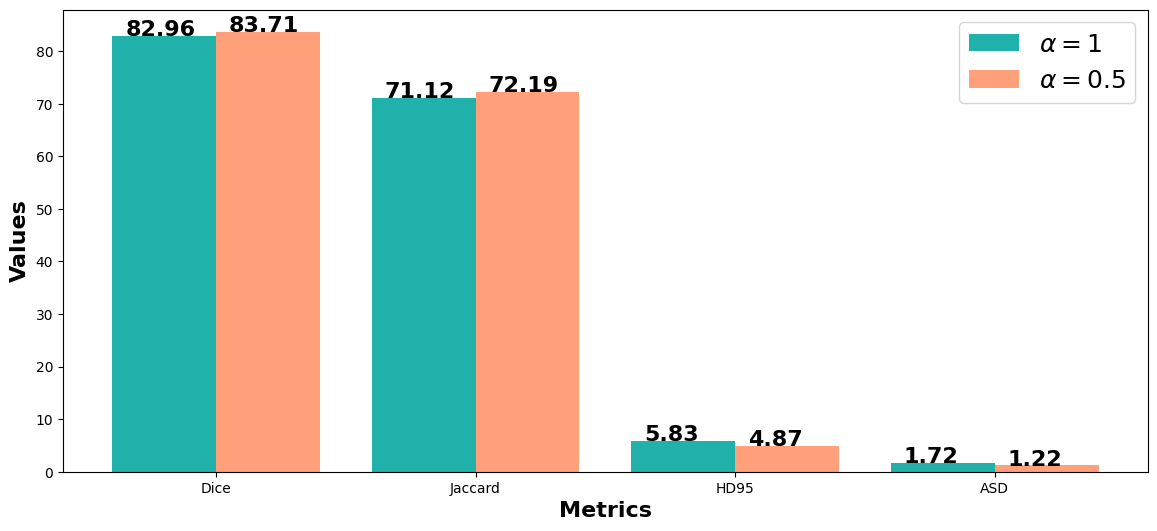}
		\centering
		\caption{Results obtained by changing   $\alpha$ values in the supervised loss function on the pancreas dataset with 20\% labeled data.}
		\label{fig:bar}
	\end{center}
\end{figure} 

\begin{figure}[h]
	\begin{center}
		\centering
		\includegraphics[width=0.45\textwidth]{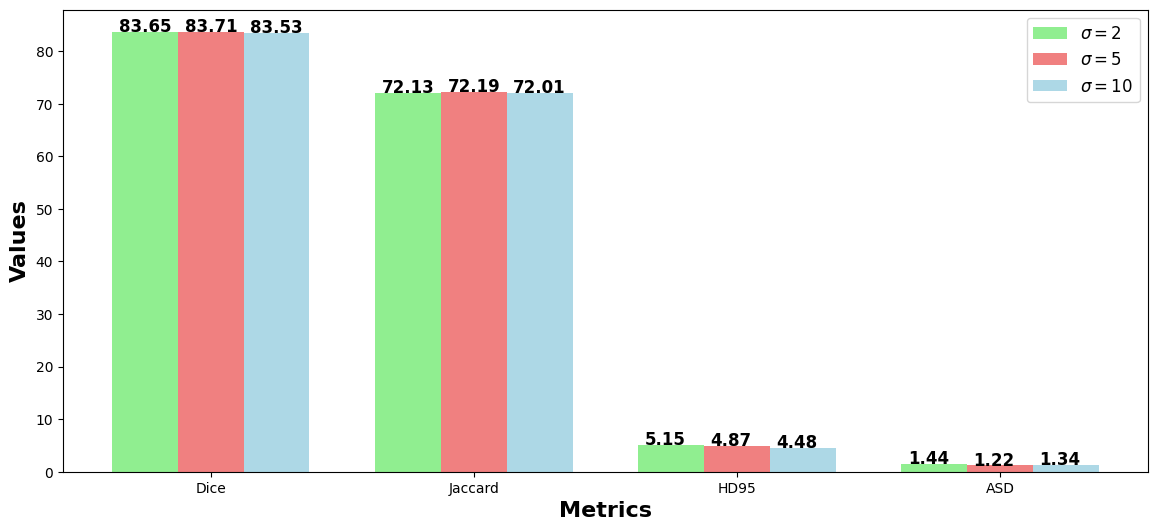}
		\centering
		\caption{Results obtained by changing parameter $\sigma$ on pancreas dataset with 20 \% labeled data.}
		\label{fig:bar1}
	\end{center}
\end{figure}

\begin{figure*}[h]
\begin{center}
\includegraphics[scale=0.7]{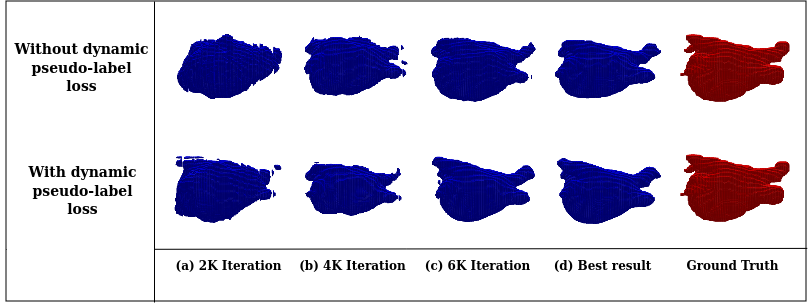}
\end{center}
\caption{Segmentation visualizations of our method w/ or w/o dynamic pseudo-label loss. The first and second rows represent segmentation results obtained at various iterations w/o and w/ dynamic pseudo label loss, respectively.}
\label{fig:LA}
\end{figure*}

\subsubsection{Introducing diversity in the model}

In this section, we discuss the significance of the dissimilarity between co-training models in determining their final performance. Our approach is influenced by the widely adopted CutMix strategy \cite{yun2019cutmix}, which has demonstrated considerable efficacy \cite{bai2023bidirectional}. We use the conventional CutMix method, where the rectangular patch is cut out from one image, and it is replaced by a patch of the same size from another randomly selected image, as shown in Figure \ref{fig:cut}. We assign the CutMix unlabeled image to only one subnet while the other subnet receives the original image. We generate a random mask $\mathcal{M}\in\{0,1\}^{W\times H\times L}$ to facilitate a cut mix between a pair of unlabeled images. Subsequently, we compute the new images as follows:

\begin{equation}\label{cutmix}
   {x}^{u_{cut}}_p = {x}^u_{p}\odot\mathcal{M}+ {x}^u_{q}\odot\left(\textbf{1} - \mathcal{M} \right)
\end{equation}

\begin{equation}\label{cutmix}
   {x}^{u_{cut}}_q = {x}^u_{q}\odot\mathcal{M}+ {x}^u_{p}\odot\left(\textbf{1} - \mathcal{M} \right)
\end{equation}

where ${x}^u_{p}$, ${x}^u_{q}\in\mathcal{D}_u$, $p\neq q$, $\textbf{1}\in\{1\}^{W\times H\times L}$, and $\odot$ means element-wise multiplication. 
This strategy enhances the diversity between the subnets and promotes more effective co-training, as shown in Table \ref{tab_ablation1}.

\section{Experiments}
\label{sec.exp}

\subsection{Dataset and Pre-Processing}
\label{sec.exp_datasets}

We evaluate our approach on three publicly available datasets: the LA dataset \cite{xiong2021global} (100 MR images), the pancreas dataset \cite{clark2013cancer} (82 CTA scans), and Brats-2019 dataset \cite{bakas2020brats}.

\subsubsection{3D Left Atrium Segmentation MR Dataset} 

A total of 100 3D gadolinium-enhanced MR images with an isotropic resolution of 0.625×0.625×0.625 comprise the Left Atrium (LA) dataset. We use the same preprocessing steps as the previous study \cite{wu2022mutual}, in which the training volumes are arbitrarily cropped to a size of $112 \times 112 \times 80$ as the model's input during training. During inference, segmentation results are obtained using a sliding window of identical dimensions, with a stride of $18 \times 18 \times 4$. We report the results for two scenarios: one employing 5\% labeled data and the other utilizing 10\% labeled data.

\subsubsection{Pancreas-CT Dataset} 
There are 82 3D abdominal contrast-enhanced CT scans in the Pancreas-CT dataset, each having a fixed resolution of 512 × 512 pixels and varying thickness ranging from 1.5 to 2.5 mm. We utilize 62 images for training following \cite{wu2022mutual} and evaluate performance on the remaining 20 scans. We crop the CT images around the pancreatic region and apply a soft tissue CT window of $[-120, 240]$ HU, in accordance with the methods of \cite{wu2022mutual}. During training, volumes are randomly cropped to dimensions of $96 \times 96 \times 96$, and during inference, a stride of $16 \times 16 \times 16$ is employed. We report the results for two scenarios: one employing 10\% labeled data and the other utilizing 20\% labeled data.

\subsubsection{Brats-2019 Dataset} Brats-2019 dataset \cite{bakas2020brats} contains MRI images from 335 glioma patients taken from various medical centers. Each patient's MRI dataset has four modalities with pixel-wise annotations: T1, T1Gd, T2, and T2-FLAIR. Based on \cite{su2024mutual}, we use 250 images for training, 25 for validation, and evaluate performance on the rest 60 scans. During training, we arbitrarily extract patches of size 96 × 96 × 96 voxels as input while employing a sliding window approach with a stride of 64 × 64 × 64 voxels for testing. We report the results for two scenarios: one employing 10\% annotated data and the other utilizing 20\% annotated data.

\subsection{Implementation Details and Evaluation Metrices}

All of our experiments were conducted using a consistent random seed on an NVIDIA A5000 GPU. We employed U-Net as the backbone for the BraTS-19 dataset and V-Net \cite{milletari2016v} for the LA and pancreas-CT datasets to ensure fairness in comparison with previous approaches. The SGD optimizer is used to train the model using an initial learning rate of $5 \times 10^{-2}$, momentum of 0.9, and weight decay factor of $10^{-4}$ for the LA and Brats-19 datasets. Similarly, for the pancreas dataset, the model is trained using the SGD optimizer, with an initial learning rate of $2.5 \times 10^{-2}$, a momentum of 0.9, and a weight decay factor of $10^{-4}$. The model is trained for 12k iterations on the LA dataset and 15k iterations on the pancreas and Brats19 datasets. With two unlabeled patches and two labeled patches, the batch size is set to 4.
Four metrics were used for evaluation: average surface distance (ASD), 95\% Hausdorff distance (95HD), Jaccard (JC), and Dice similarity coefficient (DSC). We indicated our results in bold, where our proposed method surpasses the existing methods.

\subsection{Comparison with state-of-the-art methods}

\subsubsection{Comparison on the LA Dataset} 

Table \ref{tab:LA_4} and Table \ref{tab:LA_8} show quantitative comparison results for the LA dataset. We evaluate our proposed approach against several methodologies: UA-MT \cite{yu2019uncertainty}, SASSNet \cite{li2020shape}, URPC \cite{luo2021efficient}, MC-Net \cite{wu2021semi}, DTC \cite{luo2021semi}, SS-Net \cite{wu2022exploring}, Simcvd \cite{you2022simcvd}, ACTION++ \cite{you2023action++}, BCP \cite{bai2023bidirectional}, DMD \cite{xie2023deep}, and Cross-ALD \cite{nguyen2023cross}. In line with BCP \cite{bai2023bidirectional}, we perform the experiment on the LA dataset with varied labeled ratios, i.e., 5 \% and 10 \%. To maintain fairness in comparisons, the results of other methods were obtained using the same experimental settings as outlined in \cite{bai2023bidirectional}.
As depicted in Table \ref{tab:LA_4} and Table \ref{tab:LA_8}, our method outperforms other techniques in all evaluation metrics by a significant margin. To ensure a fair comparison, we evaluate our results against methods utilizing the same data split as ours. Our method achieves notable enhancements in Dice, Jaccard, 95HD, and ASD metrics, surpassing the second-best performance by 1.53\%, 2.46\%, 2.42, and 0.16, respectively, for the 5\% setting. Similarly, for the 10\% data, we outperform the second-best approach by 1.17\%, 1.99\%, 1.85, and 0.09. These results are obtained without conducting any post-processing, ensuring fair comparisons with other methods.

\subsubsection{Comparison on the Pancreas Dataset}

Table \ref{tab:pancreas} shows quantitative comparison results for the pancreas dataset. The pancreas, situated deep within the abdomen, exhibits significant variations in size, location, and shape. Furthermore, pancreatic CT volumes exhibit a greater intricate background than MRI volumes of the left atrium. Consequently, pancreas segmentation presents greater challenges compared to left atrial segmentation. Therefore, to demonstrate the effectiveness of our method, we conduct experiments on the pancreas dataset. In line with the method \cite{wu2022mutual}, we carry out experiments on the Pancreas-NIH dataset using labeled ratios of 10\% and 20\%. We compared our approach with UA-MT \cite{yu2019uncertainty}, SASSNet \cite{li2020shape}, URPC \cite{luo2021efficient}, MC-Net+ \cite{wu2022mutual}, DTC \cite{luo2021semi}, SS-Net \cite{wu2022exploring}, SC-SSL \cite{miao2023sc}, MCCauSSL \cite{miao2023caussl}, and MLRPL \cite{su2024mutual}, as depicted in Table \ref{tab:pancreas}. To ensure a fair comparison, we evaluate our results against methods utilizing the same data split as ours. Our approach demonstrates notable enhancements in Dice, Jaccard, 95HD, and ASD metrics, surpassing the second-best method by 3.13\%, 3.92\%, 1.372, and 0.042, respectively, in the 10\% setting. Similarly, for the 20\% dataset, we outperform the second-best method by 2.18\%, 2.84\%, 1.942, and 0.112, respectively.

\subsubsection{Comparison on the BRATS-2019 Dataset}

Given the considerable variation in tumor appearance and the uncertainty surrounding tumor boundaries, brain tumor segmentation proves to be a more challenging task compared to organ segmentation. Thus, in order to showcase the effectiveness of our strategy, we perform experiments using the Brats-2019 dataset. Following \cite{su2024mutual}, we carry out experiments on the Brats-2019 dataset employing labeled ratios of 10\% and 20\%. We compared our method with DAN \cite{zhang2017deep}, CPS \cite{chen2021semi}, EM \cite{vu2019advent}, DTC \cite{luo2021semi}, CCT \cite{ouali2020semi}, URPC \cite{luo2021efficient},
CPCL \cite{xu2022all}, AC-MT \cite{xu2023ambiguity} and MLRPL \cite{su2024mutual}. 
The outcomes displayed in Table \ref{tab:brats2019} illustrate that our proposed approach achieves notable enhancements in Dice, Jaccard, 95HD, and ASD metrics (i.e., surpassing the second best by 1.42\%, 1.61\%, 1.42 and 0.1, respectively) for the 10\% setting. Same way for 20\% data, we surpass the second best by 1.18\%, 1.4\%, 0.44, and 0.3. These results are obtained without conducting any post-processing, ensuring fair comparisons with other methods.

\subsection{Ablation Study}

\subsubsection{Effects of different components}

Our method includes two losses, i.e., fix pseudo-label loss $\mathcal{L}_{fix}^u$ and dynamic pseudo-label loss $\mathcal{L}_{dyn}^u$ and another component CutMix to introduce diversity in both the models. Table \ref{tab_ablation1} illustrates the individual contributions of these losses and modules in our approach. To demonstrate the impact of each component, we initially conducted experiments using only fixed pseudo-label loss and the Cut-Mix operation. This reveals a performance improvement of over 9\% on dice metric, compared to using only supervised loss. Secondly, to assess the effect of dynamic loss, We conduct experiments using only CutMix and dynamic loss, revealing that dynamic loss alone improves Dice, Jaccard, 95HD, and ASD by 2.12\%, 2.98\%, 4.66, and 0.76, respectively, compared to the fixed pseudo-label loss. Subsequently, we perform experiments using both fixed and dynamic loss, resulting in further performance enhancement. Finally, to assess the impact of the CutMix module, we conduct an experiment where both fixed pseudo-label and dynamic pseudo-label loss are utilized, but both subnets receive the same unlabeled image. This configuration reduces the best dice score by 0.72\% percentage points.

Figure \ref{fig:pancreas} illustrates the visualization results obtained using our approach with other methods using the pancreas dataset. Our approach demonstrates a better overlap with the ground truth label and generates fewer false segmentations while providing more intricate details compared to other methods.

\subsubsection{Changing hyperparameter values in loss function}

We further conduct ablation experiments on the hyperparameter value $\beta$, which controls the contribution of the unsupervised loss for dynamic pseudo labeled, as shown in Table \ref{tab_ablation2}. A higher weight, specifically $\beta$ value 4.0, demonstrates better model performance as compared to the other values, particularly for the pancreas dataset. Likewise, we assign the value of $\beta$ as 0.1 for the LA dataset and 1.0 for the Brats-19 dataset. We also perform ablation experiments on hyperparameter value $\alpha$ corresponding to supervised loss, as shown in Figure \ref{fig:bar} and $\alpha = 0.5$ is more suitable. Therefore, We fix $\alpha$ as 0.5 for all the experiments.

\subsubsection{Changing shift parameter $\sigma$}
We examine the influence of the shift parameter $\sigma$ on the pancreas dataset, as depicted in Figure \ref{fig:bar1}. We vary $\sigma$ values over the set $\{2, 5, 10\}$ to observe the changes in performance. Based on these empirical findings, we choose $\sigma = 5$ for all our experiments in this paper.

\subsubsection{Qualitative analysis of dynamic unsupervised loss}

Figure \ref{fig:LA} shows the segmentation visualization results for the LA dataset. The first row represents when dynamic pseudo-label loss $\mathcal{L}_{dyn}^u$ is not utilized, while the second row illustrates the visualization results for the scenario when $\mathcal{L}_{dyn}^u$ is used to train the model. It is clear that segmentation results obtained with dynamic pseudo-label loss are better as compared to results obtained when using fixed pseudo-label alone.

\section{Conclusion}

This paper introduces a novel approach for semi-supervised medical image segmentation, where we utilize multiple pseudo-labels to learn from unlabeled data. Instead of relying on a single pseudo-label, which often leads to suboptimal results, we propose using both a newly proposed dynamic pseudo-label alongside the conventional fixed pseudo-label. By leveraging multiple pseudo-labels for each unlabeled image, we introduce greater diversity into the model, which in turn results in significant improvement in performance on multiple performance metrics. Our approach can help mitigate the weaknesses or limitations of traditional fixed pseudo-labels while enhancing their strengths, resulting in more robust and accurate model predictions. Our approach is a simple yet highly effective approach to improve the model’s generalization ability and segmentation accuracy. Comprehensive experiments conducted on three benchmark datasets have validated the efficacy of our proposed methodology. We also present various ablation experiments to demonstrate the efficacy of our approach.

\balance
\normalem
\bibliographystyle{IEEEtran}
\bibliography{egbib}

\end{document}